\documentclass[aps,prb,twocolumn]{revtex4} 
\usepackage{graphicx,bm,amssymb,amsmath}

\begin{document}

\title{Temperature Dependence and Screening Models in Quantum Cascade Structures}
\author{Rikard Nelander}
\email[Email: ]{rikard.nelander@fysik.lu.se}
\author{Andreas Wacker}
\affiliation{Division of Mathematical Physics, Physics Department, Lund University, Box 118, 22100 Lund, Sweden}
\date{28 August 2008, to appear in Journal of Applied Physics}

\begin{abstract}
Different screening models in quantum cascade lasers are compared by calculating the contribution of intra-subband impurity scattering to the optical linewidth as a function of temperature. We find a strong impact of impurity  scattering which is increasing substantially with temperature. A simple isotropic bulk screening model works well if the screening length is of the order or longer than the period length of the cascade structure. 
\end{abstract}

\maketitle

\section{Introduction}

The performance of terahertz (THz) quantum cascade lasers \cite{KohlerNature2002} (QCLs) has improved fast, now working up to 186 K \cite{KumarAPL2009}, and with frequencies down to 1.2~THz~\cite{WaltherAPL2007} without the use of magnetic fields. A main issue of current research is to reach higher operation temperatures \cite{WilliamsNP2007}. This requires a good understanding of the temperature dependence for all components contributing to laser operation. Often, phonon interaction is addressed when temperature-dependent effects in lasers are discussed \cite{SirtoriNP2009, CaoPE2008}. However, as previously suggested \cite{NelanderAPL2008}, there is also a strong impact on operation by screening, which strongly depends of the electron temperature.

The fast temperature degradation of THz QCLs has previously been investigated through simulation. Indjin \textit{et.al.} \cite{IndjinAPL2003} found a strongly increasing scattering rate with temperature from the upper to the lower laser state by optical phonon emission, which lowers the population inversion. This effect has also been quantified by Jirauscheck and Lugli \cite{JirauschekPSSC2008} and Cao \textit{et.al.} \cite{CaoPE2008}. One approach to remove this effect is by applying a strong static magnetic field \cite{SirtoriNP2009} and also, increasing the optical phonon energy by changing to other material systems such as gallium nitride is expected to lower this effect. However, this thermally activated phonon emission cannot explain the empirical relation that the maximum temperature is approximately limited by the photon energy \cite{WilliamsNP2007}   
\begin{equation}
k_B T_\mathrm{max} \lesssim \hbar \omega_\mathrm{las}, \label{empirical}
\end{equation}
since if this was the dominant degradation effect, long-wavelength lasers should have slightly better temperature performance. 

A second phonon effect discussed in this context is thermal backfilling, \textit{i.e.}, electrons absorb optical phonons and are reinjected to the lower laser state. This process has been theoretically shown to play a minor role \cite{JirauschekPSSC2008} and, also, the double phonon resonance did not improve the temperature performance of the device \cite{WilliamsAPL2006}. 

A simplified relation for the gain at resonance is [see \textit{e.g.} Eq.~(1) in Ref.~\onlinecite{WackerSPIE2009}] 
\begin{equation}
g_\mathrm{las} \propto \hbar  \omega_\mathrm{las} \frac{\Delta f}{\Gamma_{\rm spect}} \label{simpGain}
\end{equation} 
where $\Delta f$ is the population inversion and $\Gamma_{\rm spect}$ is the spectral width of the transition due to scattering processes. Lasing operation requires $g_\mathrm{las}$ to overcome the losses, and thus Eq.~(\ref{empirical}) is immediately satisfied if either $\Delta f \propto 1/T$, or $\Gamma_{\rm spect} \propto T$. (Of course a combination between these together with the frequency dependence of losses is more realistic for a full quantitative understanding). Most previous studies have focused on $\Delta f$, while $\Gamma_{\rm spect}$ is also expected to increase with temperature \cite{NelanderAPL2008}. Similarly, Li \textit{et.al.} \cite{LiJPD2009} found recently that even though thermally activated phonon emission strongly reduced the population inversion with temperature, life-time broadening had to be included in the model in order to correctly estimate the experimental maximum lasing temperature. In this manuscript we investigate the temperature dependence of impurity scattering at ionized dopants, which are present in all lasers considered. This scattering process is strongly affected by screening, which, as a many-body effect, is difficult to treat numerically and approximations are necessary. In this work we investigate different approximations regarding screening and their impact on the scattering strength.

\begin{figure}
\includegraphics[width=.99\columnwidth,keepaspectratio]{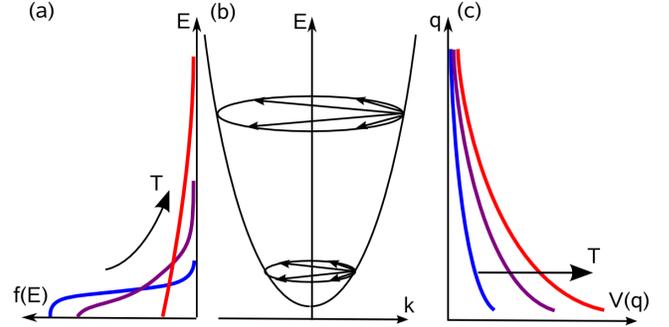}
\caption{Illustration of two, partly compensating, temperature dependent broadening effects. (a) As the temperature is increased, the electrons populate states with higher momentum. (b) Due to the constant density of states, the number of final states are constant. However, regarding elastic intra-band scattering, for large momentum, the average momentum transfer is larger than for small momentum. (c) The scattering matrix element as a function of momentum transfer. Although increasing with temperature, it is decreasing with momentum transfer.}
\label{TempEffects}
\end{figure}

The impact of temperature on this impurity scattering process is dominated by two competing processes: On the one hand, the scattering strength is increasing with temperature due to the decrease of screening. There is, however, a compensating effect: At low temperatures electrons mainly occupy the lower parts of the subband, where other momentum states at the same energy only differ by small momentum. Small momentum transfer give rise to a large impurity scattering matrix element and therefore strong scattering. At higher temperatures, electrons populate the upper part of the subbands, where the average momentum transfer is larger, giving rise to a smaller scattering rate, see Fig.~\ref{TempEffects}. The case of screening in electron-electron scattering is even more difficult due to the necessary scattering partner. 

The manuscript is organized as follows: First, we present the theory used to treat screening based on the random phase approximation. Then we show results for the scattering strength including also screening from an infinite periodic structure. After these findings we discuss the relation between the scattering strength and the actual spectral linewidth of the transition.

\section{Theory}

The method used here is based on the work of Lee and Gailbraith~\cite{LeePRB1999} regarding the screening and Ando~\cite{AndoJPSJ1985} concerning impurity scattering and broadening.

In the multi-subband structures studied here, the Dyson-like equation governing the screened Coulomb scattering matrix elements, $W_{ijkl}(q)$, is of the form,
\begin{equation} 
W_{ijkl}(q) = V_{ijkl}(q) + \sum_{m n} V_{ijnm}(q)
\Pi_{mn}(q) W_{mnkl}(q)
\label{Dyson}
\end{equation} where $V_{ijkl}(q)$ is the unscreened, or bare, Coulomb matrix element, $\Pi_{mn}(q)$ is the polarization function, $q$ is the magnitude of the in-plane momentum vector, and the indices refer to different subbands. The unscreened Coulomb matrix element is 
\begin{equation} V_{ijkl}(q) = \frac{e^2}{2 A \varepsilon_0 \varepsilon_r}
\frac{F_{ijkl}(q)}{q},
\end{equation} where $e<0$ is the electron charge, $A$ the in-plane area of the device, $\varepsilon_0$ is the vacuum permittivity, and $\varepsilon_r$ is the relative permittivity of the background material. The form factor is  
\begin{equation} 
F_{ijkl}(q) = \int \! \mathrm{d} z \!  \int \! \mathrm{d} z' \, \psi^*_i(z) \psi_j(z) \mathrm{e}^{- q | z- z'|} \psi^*_k(z') \psi_l(z').\label{Formractor}
\end{equation} Solving these double integrals can be strongly simplified by the use of the scheme presented in the latter half of Sec.~2B in Ref.~\onlinecite{BonnoJAP2005}. 

The important quantity in this context is the polarization function, where most approximations are usually done. In this work we will use the (static) random-phase approximation (RPA),
\begin{equation} 
\Pi_{nm}(\mathbf{q}) = \lim_{\delta \rightarrow 0}  2 \sum_{\mathbf{k}} \frac{f_{m,\mathbf{k} + \mathbf{q}} - f_{n,\mathbf{k}} }{E_{m,\mathbf{k} + \mathbf{q}} - E_{n,\mathbf{k}} - \mathrm{i} \delta  }
\label{RPA}
\end{equation} 
where $f_{m,\mathbf{k}}$ ($E_{m,\mathbf{k}}$) is the occupation (energy) of state $\mathbf{k}$ in subband $m$, and the factor 2 is for spin. Throughout this work we approximate the distributions $f$ with a thermalized Fermi-Dirac distribution using the total electron densities densities $n_i$ per subband obtained from our device simulation under operating conditions. If not stated otherwise we use temperature-independent $n_i$ (evaluated at 100K) and vary all band temperatures equally in order to focus on the screening effect.

The expression, similar to Eq.~(\ref{Dyson}), for determining the screened impurity scattering matrix element is 
\begin{equation} 
W_{ij}^\mathrm{imp}(q) = V_{ij}^\mathrm{imp}(q) + \sum_{m n} W_{ijnm}(q) \Pi_{mn}(q) V^\mathrm{imp}_{mn}(q),
\end{equation} where $V_{ij}^\mathrm{imp}(q)$ is the unscreened impurity scattering matrix element, 
\begin{equation} 
V_{ij}^\mathrm{imp}(q) = \frac{- e^2}{2 A \varepsilon_0 \varepsilon_r q}  \int \mathrm{d} z \, \psi_i^*(z) \psi_j(z) \mathrm{e}^{- q |z-z_\mathrm{imp}|},
\label{Vimp}
\end{equation} 
where $z_\mathrm{imp}$ is the spatial location of the impurity in the growth direction.

The linewidth of the transition at wave vector $\mathbf{k}$, is estimated by the average of the intra-band scattering rates of the two lasing states,
\begin{equation}
 \Gamma (\mathbf{k}) = \pi \sum_\mathbf{q} \left\langle
V_{uu}^2(q) + V_{ll}^2(q)  \right\rangle_\mathrm{s.c.} \delta \! \left(
E_{\mathbf{k} + \mathbf{q}} -E_\mathbf{k} \right),
\label{Eqwidth}
\end{equation} 
where $u$ and $l$ are the indices of the subbands between which the lasing transition occurs and $\left\langle ... \right\rangle_\mathrm{s.c.}$ is the average over different scattering configurations. Inter-subband scattering is weaker and neglected here. However, the temperature dependence of screening is expected to follow the same trend in both types of scattering processes. The average scattering strength related to the lasing transition is then given by
\begin{equation} 
\Gamma  = \frac{\sum_\mathbf{k} \Gamma (\mathbf{k}) |f_{u,\mathbf{k}} - f_{l,\mathbf{k}}|}{\sum_\mathbf{k} |f_{u,\mathbf{k}} - f_{l,\mathbf{k}}|}. \label{avg}
\end{equation}
In simple models, this scattering strength can be directly related to the spectral linewidth of the transition. However, this has to be taken with care as discussed in Section \ref{SecRelationStrengthWidth}.

In this work, we compare three different screening models: (i) The full RPA model [Eqs.~(\ref{Dyson},\ref{RPA})], where we include all states from $N$ periods of the QCL structure surrounding the upper and lower laser states. This is denominated by RPA-$N$. (ii) The long wavelength limit ($q \rightarrow 0)$, where
\begin{equation} 
\Pi_{ii} = - 2 \frac{A m^*}{\pi \hbar^2} f_{i,\mathbf{k}=0}, \, \, \, \Pi_{ij} =  A \frac{ n_i - n_j}{E_i - E_j} \, \, \, \text{if $i \neq j$},
\end{equation} 
with $ n_i = 2/A \sum_\mathbf{k}f_{i,\mathbf{k}=0}$, as used, {\it e.g.}, in the work of L\"{u} and Cao \cite{LuAPL2006}, and (iii) the simple isotropic screening model where the screened impurity matrix element is obtained by the replacement $q \rightarrow \sqrt{q^2 + \lambda^2}$ in the right-hand side of Eq.~(\ref{Vimp}), used by us earlier \cite{NelanderAPL2008}. $\lambda$ is the inverse screening length,  
\begin{equation} 
\lambda^2 = \frac{e^2}{\varepsilon_0 \varepsilon_r} \frac{d n_{3D}}{ d \mu}
\end{equation} 
which, in the Debye approximation (valid if $k_B T > \mu$), becomes 
\begin{equation} 
\lambda_\mathrm{Debye}^2 = \frac{ e^2 n_\mathrm{3D}}{\varepsilon_r \varepsilon_0 k_B T }, 
\label{Debye}
\end{equation} where $n_\mathrm{3D}$ is the average electron density in the structure and $\mu$ is the chemical potential of the corresponding three-dimensional electron gas.

\section{Results}

\subsection{Scattering Strength}

We focus on the THz QCL presented in Ref.~\onlinecite{KumarAPL2006} using identical parameters as in Ref.~\onlinecite{NelanderAPL2008}. Five states per period and three periods are included in the simulation.

\begin{figure} 
\includegraphics[width=1.\columnwidth,keepaspectratio]{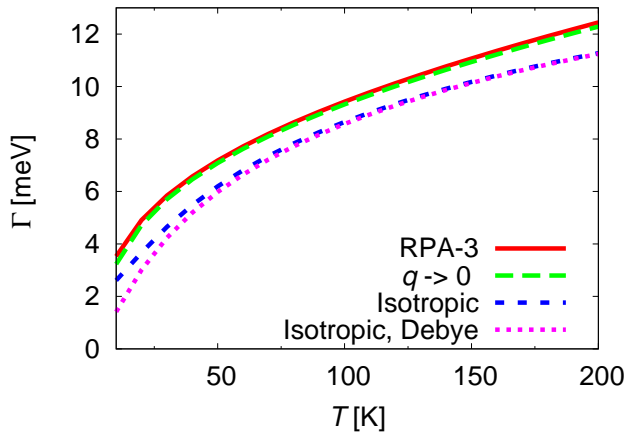}
\caption{Scattering strength due to intra-band impurity scattering for the three different screening models and the isotropic Debye approximation. For all screening models, the scattering is increasing with temperature due to less screening. Both the isotropic screening model and the long wavelength limit model slightly under-estimate the scattering (by over estimating the screening) compared to the RPA result, but are, however, excellent approximations in this situation.}
\label{widthIntraNconst}
\end{figure}

\begin{figure}
\includegraphics[width=1.\columnwidth,keepaspectratio]{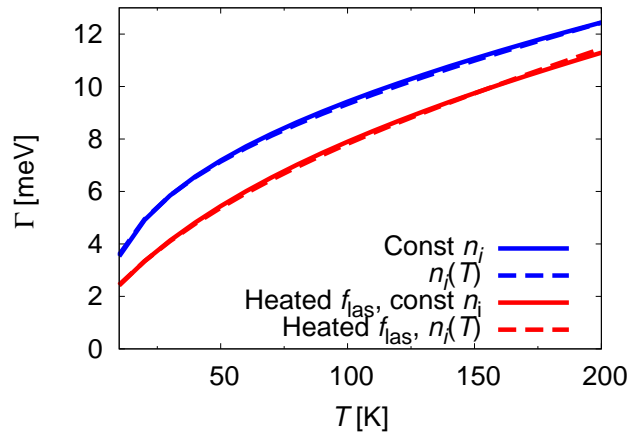}
\caption{Scattering strength calculated by RPA-3 for different choices of band temperature and densities. The temperature dependence of subband populations $n_i(T)$ hardly changes the results, while increasing the temperature of the lasing subbands by 100~K lowers the scattering strength. }
\label{widthIntraNofT}
\end{figure}

\begin{figure}
\includegraphics[width=1.\columnwidth,keepaspectratio]{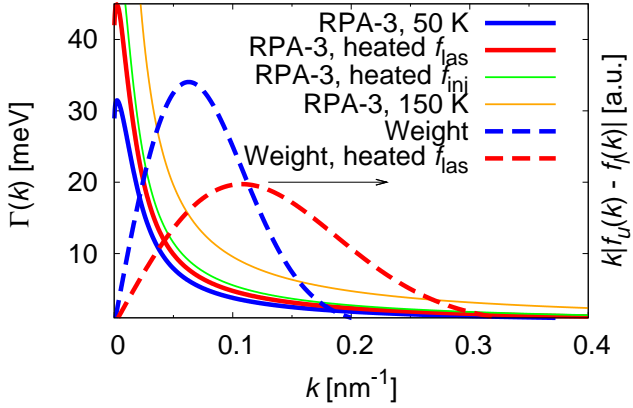}
\caption{$\Gamma (\mathbf{k})$ (full lines) and $k|f_{u,\mathbf{k}} - f_{l,\mathbf{k}}|$ (dashed lines, where the prefactor $k$ mimics the density of states in the subband). In the first case all subband temperatures are set to 50~K. In the second case the lasing subbands are 100~K warmer, and hence, the individual rates $\Gamma (\mathbf{k})$ are larger due to a slightly reduced screening. However, the electrons in lasing subbands affect the weight function $k|f_{u,\mathbf{k}} - f_{l,\mathbf{k}}|$ which is shifted to higher momentum states and therefore give a smaller average scattering strength. In contrast, if the injector subband is 100~K warmer (third case) there is even less screening and stronger scattering. This shows that the injector subband (with  59~\% of the population) is more important for screening than the lasing subbands (with 30~\% of the population). If all subbands are set to 150~K (fourth case) scattering becomes even stronger so that the average scattering strength increases taking into account the shift to higher momenta.}
\label{WidthWarmerLasingNoCorr}
\end{figure}


The scattering strength due to impurity scattering is depicted in Fig.~\ref{widthIntraNconst}. The scattering, in all approximations, is increasing with temperature due to the reduction of screening with temperature. The reduction of scattering matrix elements with momentum does not depend as strongly on temperature, as transitions with small momentum transfer are always present.

In Fig.~\ref{widthIntraNofT} the scattering strength is depicted for varying subband populations according to the results from the non-equilibrium Green's function transport simulation presented in Ref.~\onlinecite{NelanderAPL2008}. The varying subband populations have a negligible effect on the scattering strength, even though the population in, {\it e.g.}, the lower lasing subband increases by a factor of 2 over the studied temperature interval.

The electronic temperature in the heavily populated injector subband has been measured to be only slightly higher than the lattice temperature, while the less populated lasing subbands have temperatures of the order 100~K higher than the lattice \cite{VitielloAPL2005}. It is therefore of importance to calculate the scattering strength in this non-equilibrium situation. Fig.~\ref{widthIntraNofT} shows that heating of the laser level populations decreases the scattering. However, the overall trend of increasing scattering with temperature persists. As demonstrated in Fig.~\ref{WidthWarmerLasingNoCorr} the screened interaction $\Gamma(k)$ is more affected by the temperature of the injector level than that of the laser level. This shows that screening is dominated by the majority of electrons in the injector level and accordingly their temperature is the most relevant. The reduction of scattering strength by heating of the laser levels, observed in Fig.~\ref{widthIntraNofT}, is attributed to the $q$ dependence of the scattering rate, where higher values of $q$ are relevant for heated distributions.  

\subsection{Dependence on the number of periods}
\begin{figure}
\includegraphics[width=1.\columnwidth,keepaspectratio]{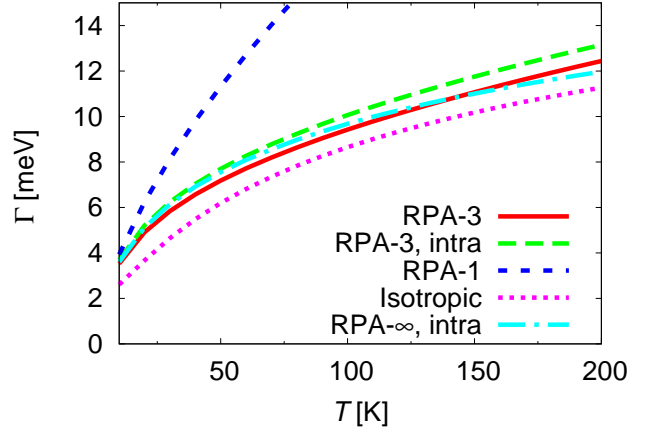}
\caption{Scattering strength for different screening models. As an estimate of the error introduced when restricting to intra-period scattering RPA-3 with and without inter-period scattering is plotted. At low temperatures, screening is efficient and all RPA models give the same results. At higher temperatures, the periodic model approaches the isotropic model.}
\label{DifferentPeriodicityNoCorr}
\end{figure}

For the data shown so far, only a finite set of periods are included in the screening models, which is a common approximation \cite{LuAPL2006,BonnoJAP2005}. For comparison, we study now the infinitely periodic system, reflecting the large number of periods in real QCL structures. First, we need to make an approximation. Up until now we have taken all combinations of states into account when studying the electron-electron matrix elements, $V_{ijkl}$. We will now neglect all matrix elements where $i$ and $j$ do not belong to the same period and/or $k$ and $l$ do not belong to the same period. This would correspond to restricting to only intra-period scattering while keeping both intra- and inter-period interaction. Taking three periods into account, this approximation has only a minor effect on the scattering strength, see Fig.~\ref{DifferentPeriodicityNoCorr}. The notation can then be changed to $V_{ijkl;h}$, where the first four indices denote subbands within a single period and $h$ is an integer denoting the distance in periods between the $ij$-pair and the $kl$-pair. $h=0$ corresponds to that all subbands are in the same period, $h=1$ to the $kl$-pair being in the first period to the right of $ij$, and so on. 

The Dyson equation, Eq.~(\ref{Dyson}), in this new notation, becomes
\begin{equation} 
W_{ijkl,h} = V_{ijkl,h} +  \sum_{m n, h'} V_{ijnm,h'}\Pi_{mn} W_{mnkl,h-h'}
\label{DysonH}
\end{equation} 
omitting the wave vector dependence. This convolution in $h$ decouples in Fourier space. Therefore, we introduce the transformation,
 \begin{equation} \begin{split} 
\tilde{V}_\kappa  &= \sum_{h}\mathrm{e}^{\mathrm {i} \kappa h}  V_h  \\ 
V_h & = \frac{1}{2 \pi} \int_{-\pi}^{\pi}
\mathrm{d} \kappa \,  \mathrm{e}^{-\mathrm {i} \kappa h} \tilde{V}_\kappa ,
\label{FT}
\end{split} \end{equation} 
where Eq.~(\ref{DysonH}) becomes
\begin{equation} 
\tilde{W}_{ijkl,\kappa} = \tilde{V}_{ijkl,\kappa} +  \sum_{mn} \tilde{V}_{ijnm,\kappa} \Pi_{mn}
\tilde{W}_{mnkl,\kappa}. \label{DysonKappa}
\end{equation} 
For large $h$, the wavefunctions of the two pairs do not overlap spatially [see Eq.~(\ref{Formractor})] and the matrix elements become particularly simple. If $H$ is the largest $h$ where the wavefunctions overlap, the matrix elements becomes $V_{ijkl, h}(q) =  \mathrm{e}^{-q (h-H) d } V_{ijkl,H}(q)$ for $h>H$ where $d$ is the length of one laser period. This exponential $h$ dependence makes the sum over $h$ in the Fourier transform, Eq.~(\ref{FT}), simple and can be easily be evaluated to infinity via a geometric sum. In computations, the symmetry $V_{ijkl;h} = V_{klij;-h}$ will be used. Eq.~(\ref{DysonKappa}) can be solved for a finite number of $\kappa$, and the screened impurity matrix element can then be calculated for infinite system.

The results can be seen in Fig.~\ref{DifferentPeriodicityNoCorr}. First, we notice that including only one period in screening strongly over-estimates scattering. Also, when including three periods, restricting to only intra-period scattering underestimates screening slightly. Finally, we see that electrons in periods further away than in the nearest neighboring periods, \textit{i.e.}, the difference between RPA-$\infty$ and RPA-3 with intra-period scattering, give a moderate effect on screening at higher temperatures.

\begin{figure}
\includegraphics[width=1.\columnwidth,keepaspectratio]{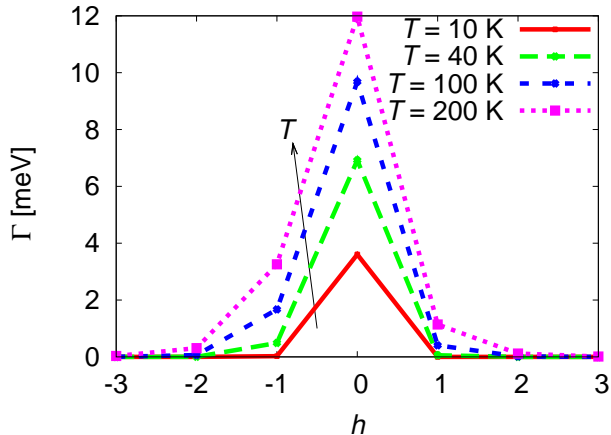}
\caption{Scattering strength due to impurity layers in neighboring periods. At low temperatures screening is efficient and scattering is spatially local. When temperature is increased, scattering at the impurities in the same period is increased, but also, scattering at impurities in neighboring periods becomes important.}
\label{DifferentPeriodicityWidthOfHNoCorr}
\end{figure}

This periodic formulation does not only give the possibility to investigate the screening contribution of electron in arbitrary periods, but also, the scattering impact on the lasing transition from impurities in any period can easily be calculated, see Fig.~\ref{DifferentPeriodicityWidthOfHNoCorr}. We see that scattering off dopants from the same periods is the strongest one, but also, at higher temperatures, dopants in neighboring periods give a substantial contribution to the scattering.

\subsection{Mid-Infrared Lasers} Up until now, only the THz device presented in Ref.~\onlinecite{KumarAPL2006} has been investigated. QCLs are divided into two different types: THz lasers which operates at frequencies below the optical phonon resonance, and mid-infrared (mid-IR) laser operating at frequencies above. Typically, mid-IR lasers are more heavily doped and, in order to reach higher optical frequencies, have more and narrower quantum wells. The period in these lasers can often clearly be divided into two spatial regions: one active region, to which the lasing subbands are confined, often only three quantum wells, and an injector region where the dopants are located. This spatial separation between the dopants and lasing subbands together with the higher electron concentration contribute to effective screening that lowers the scattering impact from impurities on the lasing transition. Also, other scattering mechanisms, such as interface roughness become much more important due to the larger conduction band offset.

\begin{figure}
\includegraphics[width=1.\columnwidth,keepaspectratio]{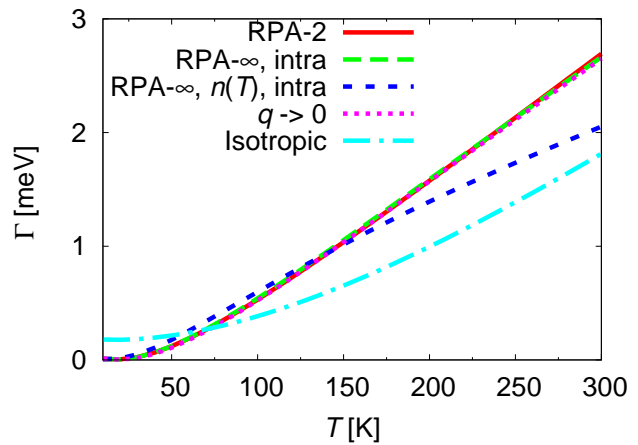}
\caption{ Scattering strength for different approximations for the mid-IR device of Ref.~\onlinecite{SirtoriAPL1998}. The curves for the RPA with 2 and infinite periods (as well as the approximation $q\to 0$) fall together, showing that the strong screening in this heavily doped structure restricts the interaction to the two neighboring periods. Changing the subband population according to a simulation based on non-equilibrium Green's functions affects the screening more in this laser compared to the THz structure due to the strong, local screening. At temperatures above 300~K the different scattering strengths increase almost linearly. Here, the RPA-$\infty$ and the isotropic screening model exhibit an almost constant difference, thus the relative error becomes smaller. The RPA-2 starts to deviate from RPA-$\infty$ around 700 K, where $1/\lambda= 22.5$~nm.} 
\label{MidIRNoCorr}
\end{figure}

\begin{figure}
\includegraphics[width=1.\columnwidth,keepaspectratio]{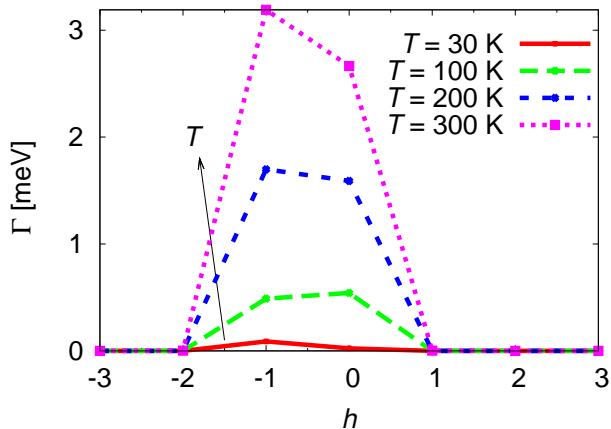}
\caption{Scattering strength due to impurity layers in neighboring periods for the mid-IR device of Ref.~\cite{SirtoriAPL1998}. As the active region where the lasing transition occurs is sandwiched between two injector regions where the dopants are located, the scattering contributions from each side is approximately equal. Due to the heavy doping, impurity layers in next neighboring periods are completely screened, also at high temperatures.}
\label{MidIRWidthH}
\end{figure}

The results from our screening model can be seen in Fig.~\ref{MidIRNoCorr} and~\ref{MidIRWidthH}, where the laser presented in Ref.~\onlinecite{SirtoriAPL1998} has been used. First, the impact of impurity scattering on the transition is much smaller for reasons mentioned above, but also, the simple isotropic screening model gives a larger error. This can be understood by comparing the screening lengths. The two lasers in this study have similar period, but the mid-IR device is much more doped and therefore has a shorter screening length, see Tab.~\ref{LasComp}. If the screening length is of the order of the laser period or longer, all electrons in the structure will contribute to screening and the isotropic screening model will be a good approximation. Then, the spatial location of the electrons does not matter since all electrons contribute to screening. In the mid-IR device, the screening length is much shorter than the period and the isotropic model screening fails. Also, the screening is more affected by temperature dependent subband populations, since the screening is strongly depending on the electron concentration in the direct vicinity of the dopants. Thus the isotropic screening model is an excellent approximation in the lightly doped THz quantum cascade lasers but is questionable for highly doped mid-IR devices, where the screening length is of the order of the individual layers.

\begin{table}
\caption{\label{LasComp} Comparison between typical THz and mid-IR devices with index $h$. Note that the high doping in the mid-IR device gives a degenerate electron gas at low temperatures resulting in an almost temperature independent screening length. In order to get a screening length in the mid-IR device of half the period length, a temperature of 720~K is required. For the THz device, the electron gas is well approximated by a non-degenerate gas resulting in Debye screening, see Eq.~(\ref{Debye}) }
\begin{ruledtabular}
\begin{tabular}{lcc} & THz Device \cite{KumarAPL2006} & Mid-IR Device
\cite{SirtoriAPL1998}\\ 
 \hline 
 $d$ & 55.4~nm & 45.3~nm\\ $n_{3D}$ & 4.06 $\times$ 10$^{15}$ cm$^{-3}$ & 8.6
$\times$ 10$^{16}$ cm$^{-3}$\\ 1/$\lambda$ (50 K) & 27.9 nm & 7.46 nm\\
1/$\lambda$ (200 K) & 55.0 nm & 10.2 nm\\
\end{tabular}
\end{ruledtabular}
\end{table}

\section{Relation between scattering strength and spectral linewidth}
\label{SecRelationStrengthWidth}

\begin{figure}[t]
\includegraphics[width=1.\columnwidth,keepaspectratio]{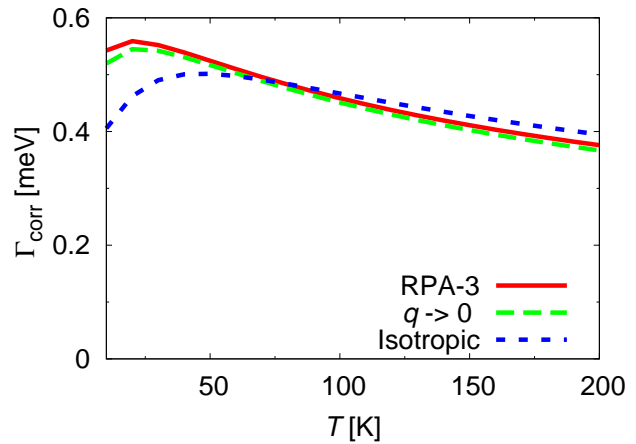}
\caption{Scattering strength due to intra-band impurity scattering including correlation effects by Eq.~(\ref{EqCorrSimple}) for the THz laser of Ref.~\onlinecite{KumarAPL2006}.}
\label{widthCorr}
\end{figure}

Calculating the spectral width of the gain transition in QCLs is quite non-trivial due to correlation effects \cite{BanitAPL2005}. A simple closed expression for the linewidth including correlation effects can be obtained by simply replacing Eq.~(\ref{Eqwidth}) by
\begin{equation}
 \Gamma_{\rm corr} (\mathbf{k}) = \pi \sum_\mathbf{q} \left\langle
|V_{uu}(q) - V_{ll}(q)|^2  \right\rangle_\mathrm{s.c.} \delta \! \left(
E_{\mathbf{k} + \mathbf{q}} -E_\mathbf{k} \right),
\label{EqCorrSimple}
\end{equation} 
see Ref.~\onlinecite{AndoJPSJ1985}. This form clearly shows the effect of correlation effects, namely that if the scattering environment of the two lasing states are similar, the optical linewidth is reduced below the sum of the lifetime induced widths of the two respective states. Also, the estimated width of the transition is now a measure of the difference in scattering strength and therefore does no longer directly reflect the scattering strength itself. The result is shown in Fig.~\ref{widthCorr} for different approximations. Again the isotropic screening model allows for a  quantitative description in comparison with the more advanced RPA-approaches. Compared to Fig.~\ref{widthIntraNconst}, the estimated width is strongly reduced and decreasing with temperature except for the low temperature regime. This difference can be understood by the fact that the impurity scattering matrix elements for $q \to 0$, which are most affected by screening, widely cancel each other in Eq.~(\ref{EqCorrSimple}), as outlined in the Appendix.

However, one has to bear in mind, that Eq.~(\ref{EqCorrSimple}) seems to under-estimate the optical linewidth compared to a full calculation, see Ref.~\onlinecite{WackerSPIE2009}. This goes in hand with the findings of Ref.~\onlinecite{NelanderAPL2008}, where a full calculation based on the isotropic, bulk screening model found an increase of the Full Width at Half Maximum from 2.7~meV at 10~K to 5.5~meV at 80~K, which is between the results of the total scattering strength from Fig.~\ref{widthIntraNconst} and the correlated scattering strength from Fig.~\ref{widthCorr}. Thus, both the expressions (\ref{Eqwidth}) and (\ref{EqCorrSimple}) should be taken with care for the determination of the width of the gain peak.

Finally, the averaging in Eq.~(\ref{avg}) is over-estimating the linewidth since the averaged gain peak has a linewidth smaller than the average linewidth. However, this effect is expected to be weak compared to the correlation effects addressed above.
\cite{AndoJPSJ1985}

\section{Conclusion}

In conclusion, the decrease of screening causes an increase of scattering strength in QCLs with temperature. If the screening length is of the order of the laser period or longer, the isotropic, bulk screening model is an excellent approximation. In this situation, which is common for typical THz QCLs, the distribution of electrons to different subbands is much  less important than the average subband temperature. For high-doped mid-IR laser structures, where the screening length is typically of the order of the layers, a microscopic RPA model is indispensable for a reliable calculation of screening. The temperature dependence of the linewidth is difficult to estimate due to correlation effect, and a full self-consistent transport calculation is needed to resolve this matter.

The authors thank C.~Weber, C.-O.~Almbladh and M.~P.~von Friesen for helpful discussions and gratefully acknowledge financial support from the Swedish Research Council (VR). 

\appendix
\section{Impact of screening on correlations in scattering matrix elements}
Screening affects the interaction over larger distances, corresponding to small wave vectors. In the limit $q \rightarrow 0$, the diagonal impurity scattering matrix element with the isotropic screening model becomes [see Eq.~(\ref{Vimp})]
\begin{equation}
V_{ii}^\mathrm{imp} \propto \frac{1}{\lambda} \int \mathrm{d} z \, \mathrm{e}^{- \lambda |z-z_\mathrm{imp}|} | \psi_i(z)|^2.
\end{equation}
The Taylor expansion of the exponential provides
\begin{equation}
V_{ii}^\mathrm{imp} =  \frac{B}{\lambda} - C_i +  O(\lambda).
\end{equation}
where $B= -e^2/2A \varepsilon_0 \varepsilon_r$, and $C_i$ are numerical constants. Since $B$ is independent of the state $i$, the square of the matrix elements including the correlation effect becomes 
\begin{equation}
|V_{ii} - V_{jj}|^2 = |C_i-C_j|^2 + O(\lambda).
\end{equation}
and without correlation effects
\begin{equation}
V_{ii}^2 + V_{jj}^2  = \frac{2B^2}{\lambda^2} + \frac{B(C_i+C_j)}{\lambda} + O(1).
\end{equation}

In the Debye approximation, $\lambda \propto 1/\sqrt{T}$, resulting in the different temperature dependence,
\begin{equation} \begin{split}
|V_{ii} - V_{jj}|^2  &\propto 1 + O(1/\sqrt{T}) \\
V_{ii}^2 + V_{jj}^2  &\propto T + O(\sqrt{T}),
\end{split} \end{equation}
clearly showing the weaker temperature dependence when including correlation effects.

\end{document}